\documentclass[11pt]{article}
\usepackage{amsmath}
\usepackage{latexsym}

\allowdisplaybreaks

\newcommand{\be}{\begin{equation}}
\newcommand{\ee}{\end{equation}}
\newcommand{\bea}{\begin{eqnarray}}
\newcommand{\eea}{\end{eqnarray}}

\newcommand{\cL}{\mathcal{L}}
\newcommand{\cO}{\mathcal{O}}

\newcommand{\I}{\mathrm{i}}
\newcommand{\vb}{\bar{v}}
\begin{document}
\begin{titlepage}
\renewcommand{\thefootnote}{\fnsymbol{footnote}}
\begin{center}
 {\LARGE\bfseries Orientifolding in $N=2$ Superspace}
\\[10mm]
\textbf{D.~Robles-Llana$^1$, M.~Ro\v{c}ek$^2$, F.~Saueressig$^1$, U.~Theis$^3$, S.~Vandoren$^1$}\\[5mm]
{\em $^1$Institute for Theoretical Physics} and {\em Spinoza
  Institute,\\ Utrecht University, Utrecht, The Netherlands}\\[3mm] 
{\em $^2$C.N.~Yang Institute for Theoretical Physics \\
Stony Brook University, NY 11794-3840, USA}\\[3mm]
  {\em $^3$Institute for Theoretical Physics \\
Friedrich-Schiller-University Jena, D--07743 Jena, Germany}\\[3mm]
\end{center}
\renewcommand{\thefootnote}{\arabic{footnote}}
\setcounter{footnote}{0}
\vspace{5mm}

\bigskip

\centerline{\bfseries Abstract} 
\medskip
\noindent
We discuss orientifold projections on superspace effective actions for
hypermultiplets. We present a simple and new mechanism that allows
one to find the K\"ahler potential and complex structure for the
$N=1$ theory directly in terms of the parent $N=2$ theory. As an
application, we demonstrate our method for Calabi-Yau orientifold
compactifications of type IIB superstrings.
\end{titlepage}
\section{Introduction}
\label{sec:intro}
Compactifications of type II strings on Calabi-Yau threefolds with
branes/fluxes lead to interesting semi-realistic models of
four-dimensional physics with $N=1$ supersymmetry (see
\cite{Grana,Blumenhagen} for recent reviews). For compact
Calabi-Yau's, cancellation of tadpoles requires the introduction
of orientifold planes. These are objects with negative tension,
and, in contrast to D-branes, have no dynamical degrees of
freedom.

The orientifolds we consider can be understood as modding out the
combined operation of worldsheet parity and an involution on the
Calabi-Yau space. In the cases we study, the
supersymmetry is reduced from $N=2$ to $N=1$. 
Here we concentrate on the effects of the projection only,
freezing brane moduli and switching off fluxes (superpotentials) and 
quantum corrections. In that case the kinetic term in the resulting 
$N=1$ supersymmetric
low-energy effective action has a very particular structure. It is
locally the product of two K\"ahler manifolds. These can be seen
as submanifolds of the special K\"ahler and quaternion-K\"ahler
moduli spaces of the $N=2$ vector and hypermultiplets
respectively.

Whereas it is straightforward to exhibit the K\"ahler structure of the $N=1$ subsector arising from truncating $N=2$ vector multiplets, the projection is more subtle for the hypermultiplet sector \cite{Andrianopoli:2001gm}: the quaternion-K\"ahler geometry of the hypermultiplet moduli space is generically not K\"ahler.
In this note we demonstrate that this problem is easily overcome by
working off-shell in $N=2$ superspace. The off-shell formulation of matter
coupled $N=2$ supergravity uses a compensator
that restores the superconformal symmetry. The orbifold projection
is then easily implemented on $N=2$ projective superfields. As a
result, the $N=2$ superconformal quotient yielding the original
$N=2$ theory reduces to an $N=1$ K\"ahler quotient which gives the
K\"ahler potential of the orientifolded theory. We illustrate our
findings by considering $\cO 3/\cO 7$ and $\cO 5/\cO 9$
orientifolds of type IIB string theory on Calabi Yau threefolds. A
similar analysis can be performed for type IIA strings.

\section{Orientifolds and projective superspace}
\label{main}

In this section we outline the orientifold projection of $N=2$
supergravity theories coupled to tensor multiplets or
hypermultiplets. As mentioned in the introduction, rather than
working with the Poincar\'e theory, we use the superconformal
tensor calculus to give an off-shell description in projective
superspace. In such a formalism the general couplings are based on
the $N=2$ superspace Lagrangian density \cite{GHR}
\begin{equation} \label{RV}
  \cL(v^I, \vb^I, x^I) = {\rm Im} \oint_C \frac{\mathrm{d}\zeta}{2\pi \I\,
  \zeta}\, H(\eta^I)\ ,
 \end{equation}
where $C$ is an appropriately chosen contour, and $\zeta$ is a
local coordinate on the Riemann sphere. The function $H(\eta)$
must be homogeneous of degree one for superconformal invariance of
the action \cite{deWit2}. Projective superfields are
defined by
\begin{equation}\label{def-eta}
\eta^I(\zeta)\equiv \frac{v^I}{\zeta}+x^I-\vb^I\zeta\ ,
\end{equation}
and describe $N=2$ tensor supermultiplets. In terms of $N=1$ superfields, the
components $v^I$ are chiral $N=1$ multiplets whereas the $x^I$ are
$N=1$ tensor multiplets containing one real scalar and a tensor.
They both have scaling weight two under dilatations, and the three
scalars form a triplet under $SU(2)_R$, which is part of the
superconformal symmetry group.

To dualize to hypermultiplets, one first defines the tensor
potential \cite{SdW}
 \begin{equation} \label{SW}
  \chi(v^I, \vb^I, x^I) \equiv - \cL +x^I \partial_{x^I} \cL\ ,
 \end{equation}
and performs a Legendre transformation by defining dual $N=1$ chiral
superfields
\begin{equation}
w_I+{\bar w}_I = \frac{\partial \cL}{\partial x^I}\ ,
\end{equation}
which have scaling weight zero. After eliminating the $x^I$ in
terms of the chiral superfields, this potential becomes the
K\"ahler potential for the superconformal hypermultiplets,
\begin{equation}
K(v,\bar v,w + \bar w)= \chi(v^I, \vb^I, x^I(v,\bar v,w + \bar
w))\ .
\end{equation}
In fact, the scalars of the hypermultiplets parametrize a hyperk\"ahler
manifold. 

Finally, to descend to the Poincar\'e theory, one couples to the
Weyl multiplet and performs a superconformal quotient. In this
formulation the physical scalar fields of the Poincar\'e theory
are given by $SU(2)_R$ and dilatation invariant combinations of
the tensor multiplet scalars.

The orientifold projections we consider break half of the
supersymmetries and truncate the theory to $N=1$. As we explicitly
demonstrate in the next section, the orientifold projection can be
implemented in projective superspace by defining a parity operator
on the complex coordinate $\zeta$, and requiring projective
superfields to be either parity-even or parity-odd
\begin{eqnarray}\label{OP}
&&\eta(-\zeta)= \eta(\zeta) \,\,\,\,\,\Longleftrightarrow v={\bar v}=0 \ ,\nonumber\\
&&\eta(-\zeta)= -\eta(\zeta) \Longleftrightarrow x=0\ .
\end{eqnarray}
We see that this projects out either the $N=1$ chiral superfield
$v$ or the $N=1$ tensor multiplet $x$, and hence there is only
$N=1$ supersymmetry left over, as required.

Since we are working in an off-shell formulation, this projection
can be carried out on the superspace Lagrangian density $\cL$ or
on the K\"ahler potential $K$. The different $N=2$ tensor
multiplets $\eta(\zeta)$ can be subject to either of the two
conditions in \eqref{OP}. The resulting $N=1$ theory is then based
on the K\"ahler potential $K(v,\bar v,w + \bar w)$ with a number
of chiral multiplets $v^M$ and a number of $N=1$ tensor multiplets
$x^u$. One then integrates this function over the standard $N=1$
superspace measure, and the theory still has $N=1$ superconformal
invariance with dilatation and $U(1)_R \subset SU(2)_R$ symmetry
inherited from the $N=2$ theory. To descend to the Poincar\'e
theory, one couples to the $N=1$ Weyl multiplet and performs an
$N=1$ superconformal quotient. This can be done by first defining
the scale and $U(1)_R$ invariant coordinates (with $M=\{0,A\}$)
\begin{equation}\label{CS}
\tau^M \equiv \frac{v^M}{v^0}=\{1,\tau^A\}\ .
\end{equation}
In other words, $v^0$ is the compensator. Gauge-fixing dilatations
and $U(1)_R$ gives old minimal $N=1$ supergravity coupled to
chiral multiplets~\footnote{There is an alternative way of performing
the superconformal quotient by using an $N=1$ tensor multiplet as
a compensator. This yields new minimal $N=1$ supergravity, but we
do not discuss this option in this paper.}. Because of the
homogeneity properties, $v^0$ can be scaled out of the K\"ahler
potential $K$ and after the $N=1$ superconformal quotient, the
resulting K\"ahler potential for the chiral multiplets is
\begin{equation}\label{K-pot}
{\cal K}(\tau^A,{\bar \tau}^A,w^u+{\bar w}^u) = - \log
K(\tau^A,{\bar \tau}^A,w^u+{\bar w}^u)\ .
\end{equation}
In more geometric terms, this quotient is just a standard $U(1)$
K\"ahler quotient \cite{Lindstrom:1983rt}.

\section{Calabi-Yau orientifolds of IIB superstrings}
\label{sec:CY}
We consider the compactification of type IIB strings on Calabi-Yau
orientifolds leading to $N=1$ supersymmetric actions in four
dimensions. After briefly summarizing the results of \cite{GL}, we
rederive them starting from the superspace formulation of the $N=2$
supersymmetric theory and apply the orientifold projection
directly in superspace.

\subsection{Orientifold projections and complex structures}

The four-dimensional effective actions have been worked out in
\cite{GL}. Before orientifolding the massless fields of the $N=2$
supergravity theory are summarized in Table \ref{T.1}.
\begin{table}[h!]
\begin{center}
\begin{tabular}{|c|c|c|}
\hline \hspace*{5mm} gravity multiplet \hspace*{5mm} & \hspace*{5mm} 1 \hspace*{5mm} & \hspace*{5mm}
$(g_{\mu \nu}, V^0 )$ \hspace*{5mm}\\ \hline vector multiplets &
$h^{(1,2)}$ & $(V^K, X^K)$ \\ \hline tensor multiplets &
$h^{(1,1)}$ & $(t^a, b^a, c^a, E_2^a)$ \\ \hline double-tensor
multiplet & 1 & $(B_2, C_2, \phi, l)$ \\ \hline
\end{tabular}
\end{center}
\caption{\label{T.1} Massless spectrum of type IIB strings
compactified on a Calabi-Yau threefold.}
\end{table}
%

There are two ways of performing the orientifold projection, which
lead to CY orientifolds with either $\cO 3 / \cO 7$ or $\cO 5 /
\cO 9$ planes. From the point of view of the $N=2$ effective
theory the difference between these theories arises from the
double tensor multiplet sector. In $\cO 3 / \cO 7$ orientifolds
the $B_2, C_2$ tensors are projected out and one is left with a
chiral multiplet containing $\phi, l$. For the $\cO 5 / \cO 9$
case the $l, B_2$ are projected out and $\phi, C_2$ form a linear
multiplet.

To find the spectrum of the other fields, one decomposes the
cohomology classes $H^{(1,1)}$ and $H^{(1,2)}$ into subspaces
which are even and odd under the orientifold action.
\be H^{1,1} = H^{(1,1)}_+ \oplus H^{(1,1)}_- \; , \qquad H^{(1,2)}
= H^{(1,2)}_+ \oplus H^{(1,2)}_- \, , \ee
with dimensions $h^{(1,1)}_+, h^{(1,1)}_-, h^{(1,2)}_+$ and
$h^{(1,2)}_-$, respectively. From this one readily works out the
$N=1$ spectrum of the orientifolded theory. The fields appearing
in the $\cO3/\cO7$ and $\cO5/\cO9$ are then summarized in Table
\ref{T.2}.
\begin{table}[h!]
\begin{center}
\begin{tabular}{|c||c|c||c|c|}
\hline & \multicolumn{2}{c||}{$\cO 3 / \cO7$} &
\multicolumn{2}{c|}{$\cO 5 / \cO9$} \\ \hline \hline
 \hspace*{5mm} gravity multiplet \hspace*{5mm} &
\hspace*{5mm} \raisebox{-0.6ex}[0ex]{1} \hspace*{5mm} &
\hspace*{5mm} \raisebox{-0.6ex}[0ex]{$g_{\mu \nu}$} \hspace*{5mm}
& \hspace*{5mm} \raisebox{-0.6ex}[0ex]{1} \hspace*{5mm} &
\hspace*{5mm} \raisebox{-0.6ex}[0ex]{$g_{\mu \nu}$} \hspace*{5mm} \\[1.5ex] \hline \hline
\raisebox{-0.8ex}[0ex]{vector multiplets} &
\raisebox{-0.8ex}[0ex]{$h^{(1,2)}_+$} &
\raisebox{-0.8ex}[0ex]{$V^k$}  &
\raisebox{-0.8ex}[0ex]{$h^{(1,2)}_-$} &
\raisebox{-0.8ex}[0ex]{$V^\kappa$} \\[1.5ex] \hline
\raisebox{-0.8ex}[0ex]{chiral multiplets} &
\raisebox{-0.8ex}[0ex]{$h^{(1,2)}_-$} & \raisebox{-0.8ex}[0ex]{$
X^\kappa$} & \raisebox{-0.8ex}[0ex]{$h^{(1,2)}_+$} &
\raisebox{-0.8ex}[0ex]{$ X^k$} \\[1.5ex]
\hline \hline
  &
\raisebox{-0.8ex}[0ex]{1} & \raisebox{-0.8ex}[0ex]{$(\phi, l)$} &
\raisebox{-0.8ex}[0ex]{$-$} &
\raisebox{-0.8ex}[0ex]{$-$} \\[1.5ex]
\raisebox{1.6ex}[-1.6ex]{chiral multiplets}  &
\raisebox{-0.8ex}[0ex]{$h^{(1,1)}_-$} &
\raisebox{-0.8ex}[0ex]{$(b^\lambda, c^\lambda)$} &
\raisebox{-0.8ex}[0ex]{$h^{(1,1)}_+$} &
\raisebox{-0.8ex}[0ex]{$(t^\sigma , c^\sigma )$} \\[1.5ex] \hline
& \raisebox{-0.8ex}[0ex]{$-$} & \raisebox{-0.8ex}[0ex]{$-$} &
\raisebox{-0.8ex}[0ex]{1} &
\raisebox{-0.8ex}[0ex]{$(\phi, C_2)$} \\[1.5ex]
\raisebox{1.6ex}[-1.6ex]{linear multiplets} &
\raisebox{-0.8ex}[0ex]{$h^{(1,1)}_+$} &
\raisebox{-0.8ex}[0ex]{$(t^\sigma, E_2^\sigma)$} &
\raisebox{-0.8ex}[0ex]{$h^{(1,1)}_-$} &
\raisebox{-0.8ex}[0ex]{$(b^\lambda, E^\lambda_2)$} \\[1.2ex] \hline
\end{tabular}
\end{center}
\caption{\label{T.2} Bosonic fields in the spectrum of Calabi-Yau
orientifold compactifications with $\cO3/\cO7$ and $\cO5/\cO9$
planes.}
\end{table}
To bring the $N=1$ effective action into standard form
one has to find a complex structure for the chiral fields spanning
a K\"ahler space. In the vector multiplet sector this is simply
the complex structure provided by the $N=2$ vector multiplet
scalars. For the remaining chiral multiplets, these holomorphic coordinates
are provided by
\be\label{CS:O3} \tau \equiv l + \I {\rm e}^{-\phi} \, , \qquad
G^\lambda \equiv c^\lambda - \tau b^\lambda \, , \qquad {\mbox
{and}} \qquad \tau^\sigma \equiv {\rm e}^{-\phi} \, t^\sigma + \I
\, c^\sigma\ , \ee
for the $\cO3/\cO7$ and $\cO5/\cO9$-orientifold, respectively. We
collectively denote these complex coordinates by $\tau^A$ and will
show that they coincide with (\ref{CS}). For the clarity of notation, we have
$\tau^A=\{\tau,G^\lambda\}$ ($\cO3/\cO7$) and
$\tau^A=\{\tau^\sigma\}$ ($\cO5/\cO9$). One can then determine the
K\"ahler potential in terms of $\tau^A$, following \cite{GL}. We give explicit formulae in later subsections.

\subsection{Superspace description}
The fields introduced above can be expressed through $SU(2)_R$ and
dilatation invariant combinations of the scalar fields of the
superconformal theory. The resulting expressions for the $N=2$
theory are given by \cite{Neitzke:2007ke,Robles-Llana:2006is}
\be\label{Poincare-coord} l + \I \, {\rm e}^{-\phi} = \frac{1}{2
\sqrt{2} (r^0)^2} \left[ \vec{r}\,^0 \cdot \vec{r}\,^1 + \I \, |
\vec{r}\,^0 \times \vec{r}\,^1 | \, \right] \, , \quad b^a + \I \, t^a
=  \frac{\eta^a_+}{\eta^1_+} \, , \quad \sqrt{2} (l b^a - c^a ) =
\frac{\vec{r}\,^0 \cdot \vec{r}\,^a}{2 (r^0)^2} \, . \ee
Here ${\vec r}\,^I = \left[ \, -i (v^I - \vb^I) \, , \, v^I + \vb^I \, , \,
x^I \, \right]$ and, with $I=\{0,\Lambda\}$ and $\Lambda=\{1,a\}$
\be \eta^\Lambda_+ = x^\Lambda + \frac{x^0}{2} \left(
\frac{v^\Lambda}{v^0} + \frac{\vb^\Lambda}{\vb^0} \right) +
\frac{r^0}{2} \left( - \frac{v^\Lambda}{v^0} +
\frac{\vb^\Lambda}{\vb^0} \right)\ . \ee
Finally, the (tree-level) tensor potential is given by
\cite{Robles-Llana:2006is}
 \begin{equation} \label{chi_class}
  \chi = \sqrt{2}\, |\vec{r}\,^0| {\rm e}^{-2\phi}\ V(t)\ ,
 \end{equation}
with $V(t)=\tfrac{1}{3!}\kappa_{abc}t^a t^b t^c$, where
$\kappa_{abc}$ are the triple intersection numbers of the
Calabi-Yau manifold. Similarly, an explicit expression for the
superspace Lagrangian density was given in \cite{Neitzke:2007ke}.

We can now apply the orientifolding on the level of the off-shell
theory. We first consider the $\cO 3 / \cO 7$ orientifold before
turning to the $\cO5 / \cO9$ case.
\subsubsection{Orientifolding to $\cO 3 / \cO 7$}
This truncation can be effected by setting the
$N=1$ linear superfields
\be \label{o3o7} \underbrace{x^0 = 0 \; , \quad x^1 = 0}_{\mbox{universal sector}} \, , \qquad \qquad \underbrace{x^\lambda = 0}_{\mbox{$h^{(1,1)}_-$ parity odd}} \, , \qquad \qquad \underbrace{ v^\sigma = 0}_{\mbox{$h^{(1,1)}_+$ parity even}} \, . \ee
Substituting this truncation into \eqref{Poincare-coord} gives
\be\label{DA-O3} \tau \equiv l + \I {\rm e}^{-\phi}= \frac{1}{2{\sqrt
2}}\frac{v^1}{v^0}\ ,\qquad G^\lambda \equiv c^\lambda - \tau b^\lambda = - \frac{1}{2
\sqrt{2}} \, \frac{v^\lambda}{v^0} \, ,\quad t^\sigma = \I |v^0| \frac{x^\sigma}{v^1 \, \vb^0 - \vb^1 \, v^0} \, ,\ee
and projects out  $B_2, C_2, t^\lambda , E_2^\lambda , b^\sigma , c^\sigma$.  Comparing to Table 2 we see that \eqref{o3o7} reproduces the spectrum of the $\cO 3/\cO 7$ orientifold projection together with the complex structure (\ref{CS}) for the chiral fields (recall that
in our notation $\tau^A=\{\tau,G^\lambda\}$).

{}From \eqref{K-pot} and \eqref{chi_class} it is easy to determine the K\"ahler potential. After scaling out $v^0$, we find, up to an
irrelevant additive constant, \be {\cal K}=-2\log[-i(\tau-{\bar
\tau})]-\log[V(t)]\, . \ee Taking into account the proper rescaling of $t^\sigma$ this agrees with \cite{GL}. We have
written the result in terms of $\tau$ and the real linear
superfields $t^\sigma$. The latter still need to be replaced in
terms of the chiral superfields through the Legendre transform. We
refrain from giving explicit formulae, as this procedure is
identical as in \cite{GL}.

\subsection{Orientifolding to $\cO 5 / \cO 9$}

As a second example we discuss orientifold compactifications with
$\cO 5 / \cO 9$-planes. In this case we set
\be\label{o5o9} \underbrace{x^0 = 0 \; , \quad v^1 = 0}_{\mbox{universal sector}} \, , \qquad \qquad \underbrace{v^\lambda = 0}_{\mbox{$h^{(1,1)}_-$ parity odd}} \, , \qquad \qquad \underbrace{ x^\sigma = 0}_{\mbox{$h^{(1,1)}_+$ parity even}} \, . \ee
Substituting \eqref{o5o9} into the expression for the $N=2$
Poincar\'e fields gives
\be\label{DS:O5} l = 0 \; , \qquad e^{-\phi} = \pm \frac{1}{4
\sqrt{2}} \, \frac{x^1}{|v^0|} \, , \qquad b^\lambda = \frac{x^\lambda}{x^1} \,,\qquad  \tau^\sigma = - \frac{i}{2 \sqrt{2}} \, \frac{v^\sigma}{v^0}\, ,\ee
while $l , b^\sigma, E_2^\sigma, c^\lambda, t^\lambda$ and one of the universal tensors are projected
out. Comparing to Table 2 we see that \eqref{o5o9} reproduces the spectrum of the $\cO 5/\cO 9$ orientifold projection together with the complex structure (\ref{CS}) for the chiral fields. The K\"ahler potential can again be obtained from \eqref{K-pot} and \eqref{chi_class}  after performing the appropriate dualization of the linear multiplets.

The general technique outlined in Section
\ref{main} should also apply to Calabi-Yau orientifolds of the
type IIA string \cite{GL2}. 

\subsubsection*{Acknowledgement}
  We thank the organizers of the RTN ForcesUniverse Network Workshop for a
  stimulating meeting. MR is supported in part by NSF grant no.\ PHY-0354776. 
  F.S.\ is supported by a European Commission
  Marie Curie Postdoctoral Fellowship under contract number
  MEIF-CT-2005-023966.\ UT was supported by the DFG within
  the priority program SPP~1096 on string theory. This work is partly 
  supported by EU contract MRTN-CT-2004-005104 and by INTAS contract 03-51-6346.



\begin{thebibliography}{[00]}

\bibitem{Grana}
  M.~Gra\~{n}a,
  {\it Flux compactifications in string theory: A comprehensive review},
  Phys.\ Rept.\  {\bf 423} (2006) 91,
  {\tt hep-th/0509003}.

\bibitem{Blumenhagen}
  R.~Blumenhagen, B.~K\"ors, D.~L\"ust and S.~Stieberger,
  {\it Four-dimensional string compactifications with D-branes, orientifolds
and fluxes},
  {\tt hep-th/0610327}.

\bibitem{Andrianopoli:2001gm}
  L.~Andrianopoli, R.~D'Auria and S.~Ferrara,
  {\it Consistent reduction of N = 2 --> N = 1 four dimensional supergravity
  coupled to matter},
  Nucl.\ Phys.\ B {\bf 628} (2002) 387
  {\tt hep-th/0112192}.

\bibitem{GHR}
S.~J.~Gates, Jr., C.~Hull and M. Ro\v{c}ek, \emph{Twisted
multiplets and new supersymmetric nonlinear sigma models.} Nucl.\
Phys.\ \textbf{B248} (1984) 157;
A.~Karlhede, U.~Lindstr\"om and M. Ro\v{c}ek,
\emph{Self-interacting tensor multiplets in N=2 superspace.}
Phys.\ Lett.\ \textbf{B147} (1984) 297.

\bibitem{deWit2} B.~de Wit, M.~Ro\v{c}ek and S.~Vandoren,
{\it Hypermultiplets, hyperk\"ahler cones and quaternion-K\"ahler
geometry}, JHEP {\bf 0102}, 039 (2001), {\tt hep-th/0101161}.

\bibitem{SdW} B.~de Wit and F.~Saueressig,
{\it Off-shell N = 2 tensor supermultiplets}, JHEP \textbf{0609},
062 (2006), {\tt hep-th/0606148}.

\bibitem{Lindstrom:1983rt}
  U.~Lindstrom and M.~Ro\v{c}ek,
  {\it Scalar tensor duality and N=1, N=2 nonlinear sigma models},
  Nucl.\ Phys.\ B {\bf 222} (1983) 285.


\bibitem{GL}
  T.~W.~Grimm and J.~Louis,
  {\it The effective action of N = 1 Calabi-Yau orientifolds},
  Nucl.\ Phys.\ B {\bf 699} (2004) 387,
  {\tt hep-th/0403067}.

\bibitem{Neitzke:2007ke}
  A.~Neitzke, B.~Pioline and S.~Vandoren,
  {\it Twistors and Black Holes},
  {\tt hep-th/0701214}.

\bibitem{Robles-Llana:2006is}
  D.~Robles-Llana, M.~Ro\v{c}ek, F.~Saueressig, U.~Theis and S.~Vandoren,
  {\it Some exact results in four-dimensional non-perturbative string theory},
  {\tt hep-th/0612027}.

\bibitem{GL2}
  T.~W.~Grimm and J.~Louis,
  {\it The effective action of type IIA Calabi-Yau orientifolds},
  Nucl.\ Phys.\ B {\bf 718} (2005) 153,
  {\tt hep-th/0412277}.



\end{thebibliography}
\end{document}